\documentclass{camera}
\usepackage{graphicx}  

\begin{document}

%
\title{Isospin effects and sensitive observables in the Fermi energy domain}

%
\author{M.Colonna$^{1}$, V.Baran$^{2}$ \and M.Di Toro$^{1,3}$}

%
\organization{
 $^1$LNS-INFN, I-95123, Catania, Italy\\
 $^2$NIPNE-HH, Bucharest and Bucharest University, Romania\\
	$^3$Physics and Astronomy Dept. University of Catania, Italy\\}
       
\maketitle

\begin{abstract}
We review recent results obtained for charge asymmetric systems at Fermi
energies. 
Observables sensitive to the isospin dependent part of nuclear interaction
are discussed, providing information on the symmetry energy behavior
below normal density.   
\end{abstract}

%
\section{Introduction}
The behavior of nuclear matter under several conditions of density and temperature
is of crucial importance for the understanding of a large variety of phenomena,   
ranging from the structure of nuclei and their decay modes, up to the life and the
properties of massive stars. Different mechanisms involving nuclear processes at  
fundamental level can be linked by the concept of nuclear Equation of State (EOS). 
In particular, over the past years, many efforts have been devoted to the investigation of the isovector
part of the nuclear interaction, aiming at constraining the density dependence of the
symmetry energy $E_{sym}$ (Iso-EOS) \cite{baranPR,baoPR08}. 
Transient states of nuclear matter far from normal conditions can be created
in heavy ion collisions at Fermi and intermediate energies.   
Hence reactions with neutron-rich/poor  systems and, in perspective, with exotic beams can be seen 
as a suitable tool to explore this issue.
Our strategy to constrain the nuclear interaction is to implement effective 
density functionals into transport codes, devoted
to simulate the collisional dynamics (here we follow the Stochastic Mean Field (SMF)
approach \cite{chomazPR}). Then predictions can be compared to experimental
data for specific reaction mechanisms, and related observables, particularly sensitive to
isospin effects.  We will focus on collisions
at Fermi energies, where one essentially explores the low-density 
behavior of $E_{sym}$. We will test  an asy-soft parametrization,
with an almost flat behavior below
 $\rho_0$, or an asy-stiff behavior, 
with a faster decrease at lower densities \cite{baran02}. 

\section{Isospin diffusion in semi-peripheral reactions}

We consider semi-peripheral reactions
between two nuclei (denoted by H and L) having different N/Z and we investigate the diffusion
of the initial isospin gradient.
This process involves nucleon exchange 
through the low density neck region and hence it is sensitive to
the low density behavior of $E_{sym}$ \cite{tsang92,isotr07,sherry}.
Within a first order approximation of the transport dynamics, the relaxation
of a given observable $x$ towards its equilibrium value can be expressed as:
$x_{P,T}(t) - x^{eq} = (x^{P,T} -  x^{eq})~e^{-t/\tau}$,
where $x^{P,T}$ is the $x$ value for the projectile (P) or the target (T) before the diffusion 
takes place, 
$x_{eq} = (x^P + x^T)/2$ is the full equilibrium value, $t$ is the elapsed time
and $\tau$ is the relaxation time, that depends on the mechanism under study.   
The degree of isospin equilibration reached in the collision can be inferred by looking 
at isospin dependent observables in the exit channel, such as the asymmetry $\beta$ of
projectile-like (PLF) and target-like (TLF) fragments.
It is rather convenient to construct the so-called imbalance ratio \cite{tsang92}:
\begin{equation}
R^x_{P,T} = {(x_{P,T}-x^{eq})} / {|x^{P,T}-x^{eq}|} 
\end{equation}
 Clearly, this observable measures the difference between the actual asymmetry of 
PLF (or TLF) and the full equilibrium value, normalized to the initial
distance (i.e. to the conditions before the diffusion process has started).
In the calculations the latter can be evaluated by looking at the asymmetries
of PLF ( or TLF), as obtained in the symmetric reactions HH and LL (where diffusion
does not take place), after
fast particle emission is over.
Within our approximation, the imbalance ratio simply reads: $R_{P,T} = \pm e^{-t/\tau}$.  
According to this  expression, the observable $R_{P,T}$ is actually
independent of the initial asymmetry distance between the reaction partners
and isolates the effects of the isodiffusion mechanism, whose strength  is determined by the 
relaxation time $\tau$, related to the symmetry energy.
However the degree of equilibration reached in the reaction crucially depends also
on the  contact time $t$, i.e. on the reaction centrality. 
To pin down the information on $E_{sym}$
we study the correlations between isospin equilibration and
the kinetic energy loss, that is adopted as a centrality selector \cite{soul04,isotr07}.
The simple arguments developed above are confirmed by the simulations of
$^{124}$Sn (H) +$^{112}$Sn (L) collisions at 35 and 50 MeV/u, that have been carried out
empolying the SMF treatment with  momentum dependent (MD) and momentum independent (MI) interactions 
\cite{isotr07}. 
\begin{figure}[h]
\includegraphics[width=16pc]{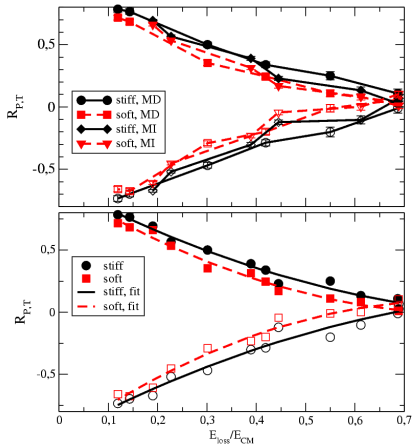}\hspace{2pc}%
\begin{minipage}[b]{12pc}\caption{\label{imb_eloss}
Imbalance ratios as a function of relative energy loss. 
Upper panel: separately for 
stiff (solid) and soft (dashed) Iso-EOS, and for 
two parametrizations of the isoscalar part of the interaction: 
MD 
(circles and squares) and MI 
(diamonds and triangles), in the projectile region (full symbols)
 and the target region 
(open symbols).
Lower panel: quadratic fit to all points for the stiff (solid), resp.
 soft (dashed) 
Iso-EOS.}
\end{minipage}
\end{figure}  
In fig.\ref{imb_eloss} we report the correlation between
$R_{P,T}$ 
and the total kinetic energy loss, normalized to the total energy available
in the center of mass $E_{cm}$, 
for the full set of calculations performed.  
On the bottom part of the figure
one can see
that all the points essentially follow a given line,
depending only on the symmetry energy parametrization adopted. A larger
equilibration (smaller $R$) is observed in the asy-soft case, corresponding to
the larger value of $E_{sym}$ at low density.
We mention that,
according to its definition, the imbalance ratio does not change if one considers
as observable $x$, instead of the asymmetry of PLF and TLF,
other observables that are linearly correlated to it and more accessible from the experimental
point of view, such as isoscaling coefficients, ratios of 
production of light isobars \cite{wci_betty} or isotopic
content of light particle emission \cite{Emma}.

An experimental study of isospin diffusion as a function of the dissipated
kinetic energy  has been recently performed by the Indra collaboration \cite{Emma}. 
Two systems, with the same projectile, $^{58}$Ni,
and two different targets ($^{58}$Ni and $^{197}$Au), at incident energies of 52 MeV/u and
74 MeV/u have been considered, with the aim to study isospin diffusion between Ni and
Au nuclei.  
An isospin-dependent variable, correlated
to the PLF asymmetry, is constructed 
from the isotopically identified light particles
emitted from the PLF, namely their average isotopic content, $(N/Z)_{CP}$.
%
The analysis of this observable in terms of imbalance ratios is not possible for this
set of reactions. Hence, 
in fig. 2 we plot directly  the results of the SMF simulations 
concerning the variable $(N/Z)_{CP}$  (lines), calculated
after de-exciting the hot primary PLF's with the help of the SIMON code \cite{Simon}. 
One can notice (see top-left panel) that isospin diffusion effects are larger in asy-soft case
(full line), as expected. For the symmetric Ni + Ni system, isospin effects are only due to
pre-equilibrium emission. 
Concerning the experimental data, 
open points show the values obtained forward in the nucleon-nucleon (NN) frame. 
In this case mid-rapidity particles and those coming from the
PLF de-excitation are mixed up. 
     The close points in fig.2 are related to the values of $(N/Z)_{CP}$ forward in the
PLF frame.
They are more representative of the isotopic content of the particles emitted
from the PLF and can be compared with 
the results of the simulations. 
When looking globally at the results for the four cases treated
here, the agreement is better when the asy-stiff EOS is used. However
for Ni+Au at 52 MeV/u, where isospin transport effects are dominant, the
close points lie in between the simulated results with the two iso-EOS. 
To conclude, 
this analysis points to a symmetry energy behavior in between the two
adopted parametrizations, in agreement with other recent estimates \cite{bettynew,cardella}.

\begin{figure}[t]
\centering
\resizebox{0.6\textwidth}{!}{%
\includegraphics{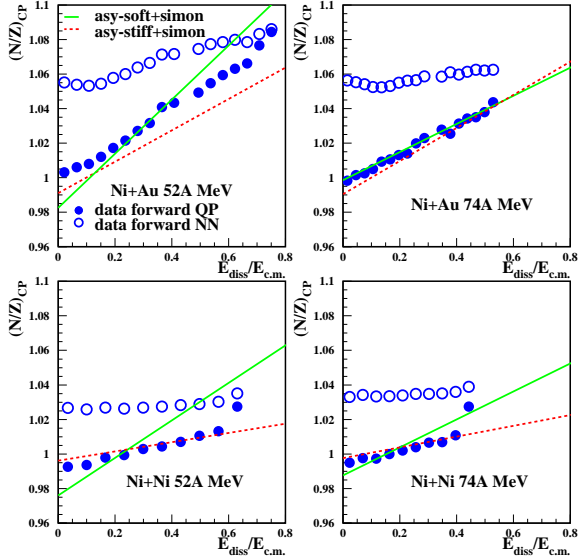}}
\caption{Isospin ratio of complex particles for Ni
 quasi-projectile vs dissipated kinetic energy, for the two reactions and the two energies.
Circles correspond to the experimental data, open for data forward of the 
N-N velocity, full for data forward in the PLF frame.
Dotted lines are for asy-soft calculations and full lines for asy-stiff. 
Adapted from ref.\cite{Emma}.}
\label{figure51} 
\end{figure}

 
\section{Isospin dynamics in neck fragmentation}

\begin{figure}[t]
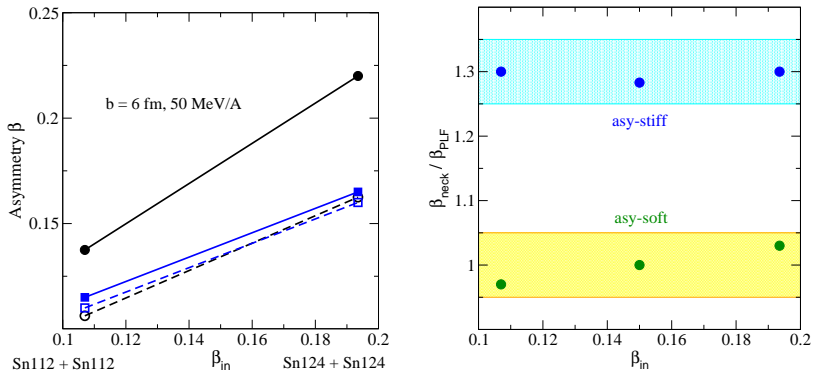

\centering
 \includegraphics[width=12pc]{fig03_a.eps}
\hskip 0.5cm
 \includegraphics[width=12pc]{fig03_b.eps}
\caption{
Left panel: asymmetry of IMF's (circles) and PLF-TLF (squares), as a function of the
system initial asymmetry, for two Iso-EOS choices: Asy-stiff (full lines) and
asy-soft (dashed lines).  
Right panel: 
Ratio between the neck IMF and the PLF asymmetries, as a function of the system
initial asymmetry. The bands indicate the uncertainty in the calculations.}
\label{nzphi}
\end{figure}

It is now quite well established that the largest part of the reaction
cross section for dissipative collisions at Fermi energies goes
through the {\it Neck Fragmentation} channel, with intermediate mass fragments
(IMF) directly
produced in the interacting zone in semiperipheral collisions on very short
time scales \cite{wcineck}. It is possible to predict interesting 
isospin transport 
effects for this 
fragmentation mechanism since clusters are formed still in a dilute
asymmetric matter but always in contact with the regions of the
projectile-like and target-like remnants almost at normal densities.
In presence of density gradients the isospin transport
is mainly ruled by drift coefficients and so
we expect a larger neutron flow to
 the neck clusters for a stiffer symmetry energy around saturation  
\cite{baranPR}.
This is shown in fig.\ref{nzphi} (left), where the asymmetry of the neck region 
is plotted for the two Iso-EOS choices, compared to
the PLF-TLF asymmetry, for  Sn + Sn reactions at 50 MeV/u, b = 6 fm.     
In order to build observables less affected by secondary decay effects, 
in fig.3 (right)  we consider the ratio of the asymmetries of the IMF's to those of the
residues for stiff and soft iso-EOS. 
This quantity 
can be estimated
on the basis of simple energy balance considerations.
Starting from a residue of mass $A_{res}$ and a neck of mass $A_{IMF}$
we assume that the mass $A$ participating in the isospin exchange
is approximately equal to the mass of the neck, while it is small relative to
the mass of the residue.  This will lead to the asymmetry  $(\beta + \Delta\beta)$
of the neck, and to a total asymmetry $\beta_{res} = [\beta(A_{res} - A) + (\beta - \Delta\beta)A]/A_{res} =
\beta - \Delta\beta A/A_{res}$ of the residue, with $\Delta\beta$ to be determined by minimization of
the symmetry energy. 
The corresponding variation of the symmetry energy is equal (apart from a constant) to:  
\begin{equation}
\Delta E_{sym} = A_{res} E_{sym}(\rho_R)(\beta-\Delta\beta A/A_{res})^2
+ A E_{sym}(\rho_I)(\beta+\Delta\beta)^2 ,
\end{equation}
where $\rho_R$ and $\rho_I$ are the densities of the residue and
neck regions, respectively.  The minimum of the variation of 
$\Delta E_{sym}$ yields 
\begin{equation}
\frac{\beta_{IMF}}{\beta_{res}} 
= \frac{E_{sym}(\rho_R)}{E_{sym}(\rho_I)} 
\end{equation}
From this simple argument the ratio between the IMF and residue asymmetries should
depend only on symmetry energy properties and, in particular, on the difference of the 
symmetry energy between the residue and the neck regions, as appropriate 
for isospin migration.  
It should also be larger than one, more so for the asy-stiff than
for the asy-soft EOS.
It is seen indeed in fig.3, that this ratio 
is nicely dependent on the iso-EOS only (being larger in the asy-stiff case) 
and not on the system considered.
If final asymmetries were affected in the same way by secondary evaporation 
in the case of neck and PLF fragments, then one could directly compare the
results of fig.3 (right) to data.  However, due to the different size and
temperature of the neck region with respect to PLF or TLF sources, 
de-excitation effects should be carefully checked with the help of
suitable decay codes. 

\section{Conclusions}

We have reviewed some aspects of the phenomenology associated with nuclear 
reactions at Fermi energies, from which new hints
are emerging to constrain the EOS of asymmetric matter below normal density.   
A considerable amount of work has already been done in this domain. 
We note also recent confirmations from structure data,
see for instance the study of 
monopole resonances in Sn-isotopes \cite{garg_prl07}.    
In the near future, thanks to the availability of both stable and rare
isotope beams, more selective analyses, also based on new exclusive 
observables, 
are expected to provide further stringent constraints.

\section{Acknowledgements}
We warmly thank H.Wolter, M.Zielinska-Pfabe, B.Borderie, E.Galichet and M.F.Rivet for inspiring discussions.





%

\end{document}